\documentclass[twoside,twocolumn,9pt]{article}
\usepackage{extsizes}
\usepackage[super,sort&compress,comma]{natbib} 
\usepackage[version=3]{mhchem}
\usepackage[left=1.5cm, right=1.5cm, top=1.785cm, bottom=2.0cm]{geometry}
\usepackage{balance}
\usepackage{widetext}
\usepackage{times,mathptmx}
\usepackage{sectsty}
\usepackage{graphicx} 
\usepackage{lastpage}
\usepackage[format=plain,justification=raggedright,singlelinecheck=false,font={stretch=1.125,small,sf},labelfont=bf,labelsep=space]{caption}
\usepackage{float}
\usepackage{fancyhdr}
\usepackage{fnpos}
\usepackage[english]{babel}
\usepackage{array}
\usepackage{droidsans}
\usepackage{charter}
\usepackage[T1]{fontenc}
\usepackage[usenames,dvipsnames]{xcolor}
\usepackage{setspace}
\usepackage[compact]{titlesec}
\usepackage{gensymb}
\usepackage{stfloats}
\usepackage{siunitx}
\usepackage{float}

\usepackage{epstopdf}
\usepackage{hyperref}

\definecolor{cream}{RGB}{222,217,201}

\begin{document}

\pagestyle{fancy}
\thispagestyle{plain}
\fancypagestyle{plain}{

\renewcommand{\headrulewidth}{0pt}
}

\makeFNbottom
\makeatletter
\renewcommand\LARGE{\@setfontsize\LARGE{15pt}{17}}
\renewcommand\Large{\@setfontsize\Large{12pt}{14}}
\renewcommand\large{\@setfontsize\large{10pt}{12}}
\renewcommand\footnotesize{\@setfontsize\footnotesize{7pt}{10}}
\makeatother

\renewcommand{\thefootnote}{\fnsymbol{footnote}}
\renewcommand\footnoterule{\vspace*{1pt}%
\color{cream}\hrule width 3.5in height 0.4pt \color{black}\vspace*{5pt}} 
\setcounter{secnumdepth}{5}

\makeatletter 
\renewcommand\@biblabel[1]{#1}            
\renewcommand\@makefntext[1]%
{\noindent\makebox[0pt][r]{\@thefnmark\,}#1}
\makeatother 
\renewcommand{\figurename}{\small{Fig.}~}
\sectionfont{\sffamily\Large}
\subsectionfont{\normalsize}
\subsubsectionfont{\bf}
\setstretch{1.125} 
\setlength{\skip\footins}{0.8cm}
\setlength{\footnotesep}{0.25cm}
\setlength{\jot}{10pt}
\titlespacing*{\section}{0pt}{4pt}{4pt}
\titlespacing*{\subsection}{0pt}{15pt}{1pt}

\fancyfoot{}
\fancyfoot[RO]{\footnotesize{\sffamily{1--\pageref{LastPage} ~\textbar  \hspace{2pt}\thepage}}}
\fancyfoot[LE]{\footnotesize{\sffamily{\thepage~\textbar\hspace{3.45cm} 1--\pageref{LastPage}}}}
\fancyhead{}
\renewcommand{\headrulewidth}{0pt} 
\renewcommand{\footrulewidth}{0pt}
\setlength{\arrayrulewidth}{1pt}
\setlength{\columnsep}{6.5mm}
\setlength\bibsep{1pt}

\makeatletter 
\newlength{\figrulesep} 
\setlength{\figrulesep}{0.5\textfloatsep} 

\newcommand{\topfigrule}{\vspace*{-1pt}%
\noindent{\color{cream}\rule[-\figrulesep]{\columnwidth}{1.5pt}} }

\newcommand{\botfigrule}{\vspace*{-2pt}%
\noindent{\color{cream}\rule[\figrulesep]{\columnwidth}{1.5pt}} }

\newcommand{\dblfigrule}{\vspace*{-1pt}%
\noindent{\color{cream}\rule[-\figrulesep]{\textwidth}{1.5pt}} }

\makeatother

\twocolumn[
  \begin{@twocolumnfalse}
\vspace{3cm}
\sffamily
\begin{tabular}{m{4.5cm} p{13.5cm} }

& \noindent\LARGE{\textbf{The role of charge and proton transfer in fragmentation of hydrogen-bonded nanosystems: the breakup of ammonia clusters upon single photon multi-ionization.}} \\
\vspace{0.3cm} & \vspace{0.3cm} \\

 & \noindent\large{Bart Oostenrijk\textit{$^{a}$}, Noelle Walsh\textit{$^{b}$}, Joakim Laksman\textit{$^{c}$}, Erik P. M\aa nsson\textit{$^{d}$}, Christian Grunewald\textit{$^{e}$}, Stacey Sorensen\textit{$^{a}$} and Mathieu Gisselbrecht\textit{$^{a}$}} \\

& \noindent\normalsize{
The charge and proton dynamics in hydrogen-bonded networks are investigated using ammonia as a model system.
The fragmentation dynamics of medium-sized clusters (1-2 nm) upon single photon multi-ionization is studied, by analyzing the momenta of small ionic fragments.
The observed fragmentation pattern of the doubly- and triply- charged clusters reveals a spatial anisotropy of emission between fragments (back-to-back). Protonated fragments exhibit a distinct kinematic correlation, indicating a delay between ionization and fragmentation (fission).
The different kinematics observed for channels containing protonated and unprotonated species provides possible insights into the prime mechanisms of charge and proton transfer, as well as proton hopping, in such a nanoscale system.}

\end{tabular}

 \end{@twocolumnfalse} \vspace{0.6cm}
  ]

\renewcommand*\rmdefault{bch}\normalfont\upshape
\rmfamily
\section*{}
\vspace{-1cm}

\footnotetext{\textit{$^{a}$~ Division of Synchrotron Radiation Research, Department of Physics, Lund University, Box 118, 22100 Lund, Sweden. E-mail: bart.oostenrijk@sljus.lu.se}}
\footnotetext{\textit{$^{b}$~ MAXIV Laboratory, Lund University, Lund, Sweden. }}
\footnotetext{\textit{$^{c}$~ European XFEL, Schenefeld, Germany.}}
\footnotetext{\textit{$^{d}$~ Institute of photonics and nanotechnology-CNR, Polytechnical University of Milan, Milan, Italy.}} 
\footnotetext{\textit{$^{e}$~ Institut f\"{u}r Chemie und Biochemie, Freie Universit\"{a}t Berlin, Berlin, Germany.}}




\section*{Introduction}

Increasing our understanding of the charge and proton transfer mechanisms in hydrogen-bonded networks has been of great importance for research in physics, chemistry and biology for decades. \cite{Nagle_1978, Staib_1995, Mazucca_2017, Marx_2006, Voth_2006} For example, these mechanisms are central in electrolyte solutions used to optimize fuel cells for energy transformation, \cite{Kreuer_2004} in acidified clouds where a modified reactivity of aerosols can change the nucleation mechanism,\cite{enghoff_2008, Li_2015} and even in biological membranes where a difference of potential can activate a `proton pump' through a membrane. \cite{DeCoursey2008, DeCoursey_2010}
A large range of molecular clusters, from an isolated molecule to a liquid, \cite{Hertel_2006} have been used as model systems for following the time evolution of chemico-physical processes. However, tracking the fundamental mechanisms becomes increasingly difficult for larger cluster systems (e.g. >1nm), which are of interest as model systems in the understanding of charge and proton transfer processes in liquids.

Synchrotron radiation in the soft X-ray range is a powerful tool for investigating these clusters, since it allows for element specific and site-selective excitation and ionization.\cite{bjorneholm09, Geiger_2002, Ruhl_2003} Photo-absorption and -ionization studies have also proven useful in providing information on the mechanisms of charge solvation, as well as on the electronic properties of various ionic species.\cite{Lindblad_2008,Ohrwall_2005} The advances in spectroscopic methods based on the multi-coincident detection of particles have opened up new avenues for studying multiply charged clusters, leading for instance to the discovery of inter-molecular decay (ICD), a mechanism that transfers the electronic excess energy between neighboring molecules.\cite{Jahnke2010,Mucke2010}

\begin{figure}[!t]
\centerline{\includegraphics[width=0.5\textwidth]{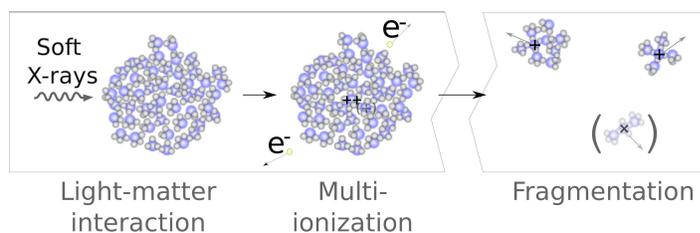}}
\caption{A schematic representation of the measurement principle. The response of ammonia clusters to multi-ionization by single soft X-rays is studied by an analysis of the fragment mass, momentum and energy, providing an insight into charge and proton transfer mechanisms.}
\label{fig:Concept}
\end{figure}

In this work, we study for the first time the fragmentation dynamics of multiply-charged ammonia clusters combining coincidence and 3D momentum imaging techniques with synchrotron radiation. The clusters are multi-ionized, and by studying the kinematics of the fragmentation, we gain information on the ultrafast dynamics of charge/proton transfer in clusters of medium size, as illustrated in Figure \ref{fig:Concept}.

\begin{figure*}[!ht]
\centerline{\includegraphics[width=0.8\textwidth]{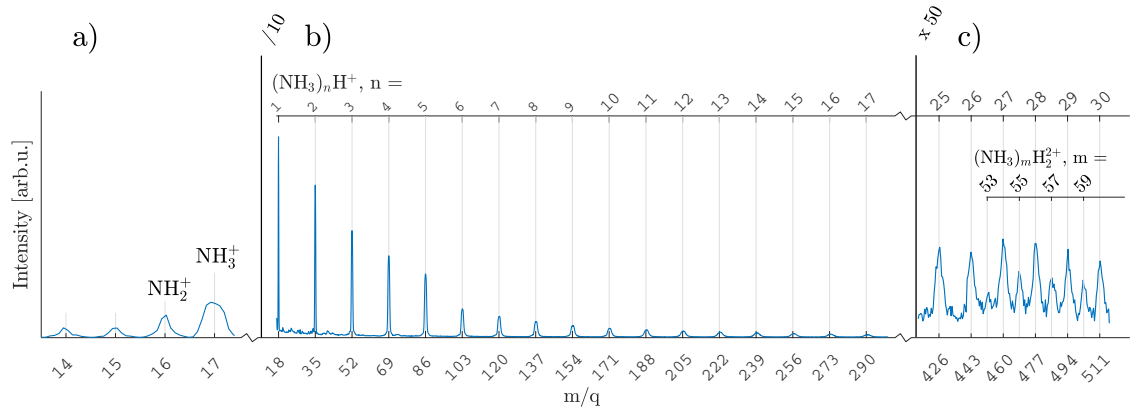}}
\caption{Mass-to-charge spectrum of charged fragments upon ionization of \ce{NH_3} clusters, separated in three regions: (a) the unprotonated cluster fragments, in coincidence with protonated clusters, as shown in (b), \ce{(NH_3)_nH^+}, and (c) the observation of the stable doubly protonated clusters around $m = 53$. Spectrum constructed from merged measurements recorded at a photon energy of 420 eV and nozzle pressures of 0.2 and 0.3 bar.}
\label{fig:1D_m2q}
\end{figure*}

Pioneering work has established that charge transfer along the hydrogen bond forms an effective path along which a proton (charge) can `hop' between molecular units, following the reaction:\cite{Cook_1979, Ceyer_1979, Shinohara_1984, Echt_1985}
\begin{align}
 \ce{NH3+-NH3-(NH3)_m -> NH2 + NH4+-(NH3)_{m}}
\end{align}
This transfer is exothermic (about 0,74 eV\cite{Ceyer_1979, Echt_1985}) and happens on a femtosecond timescale.\cite{Tachikawa2015, Farmanara_1999, Tomoda1986}
The deprotonated fragment often evaporates swiftly from the cluster, but a small fraction can remain in the same cluster as the protonated fragment.\cite{Shinohara_1985, Belau_2007}
The proton transfer generally occurs for all charge states, and the different ionic clusters act as separate entities in the neutral cluster host.\cite{COOLBAUGH198919, Peifer_1989_triple}
Upon this proton transfer, the cluster rearranges from its neutral configuration \cite{Beu_2001, GREER_1989} to shell-like structures around the ammonium (NH$_4^+$) \cite{Nakai2000201, Park_2000} in the charged state. The ammonium cation bonds to four neutral molecules as a donor, making the pentamer configuration a stable closed-shell configuration (`magic number' of $n=5$ \cite{Peifer_1989, Shinohara_1984}).
In hydrogen-bonded systems, protons can `hop' to neighboring molecules along a hydrogen bond (Grotthuss mechanism \cite{Grotthuss_1806, agmon1995, Marx_2006}), upon which the newly protonated molecule can release another proton along its hydrogen bond. 
The precise description of the diffusion mechanism in bulk liquid is difficult, since the motion involves the rearrangement of hydrogen bonds around the hopping proton.\cite{Marx_2006, Markovitch_2008}
The path of unidirectional hydrogen bonds along which the protons transfer (`proton wire') is mostly studied in aqueous systems.\cite{agmon1995, Marx_2006} However, recent studies on ammonia-complexes show that proton hopping occurs in a hydrogen-bonded model system of ammonia as well.\cite{Manca_2015}

Traditionally, multiply charged clusters are described within the framework of the Liquid Droplet Model (LDM), first proposed by Lord Rayleigh.\cite{Rayleigh_1882} According to this model, a cluster is stable beyond a critical size, i.e. when the cohesive surface tension forces overcome the charge-to-charge repulsive Coulomb forces. The critical sizes are 51 and 121 for doubly \cite{Shukla1984, COOLBAUGH198919} and triply \cite{Kreisle_1986} charged protonated ammonia clusters, respectively. Such stable clusters allows the study of (proton) chemistry within clusters.\cite{COOLBAUGH198919, Peifer_1989,Peifer_1989_triple} For unstable clusters, it was recently shown on van der Waals clusters that the fissility parameter ($X$) leaves a footprint in the correlation of fragment momenta measured.\cite{Last_2002, Hoener_2008} That is, if $X>1$, the cluster fragments through Coulomb explosion with uncorrelated momentum distributions, while close to $X \simeq 1$ the fragmentation is characterized by fission over an energy barrier \cite{Casero_1988} with strongly correlated momentum distributions. Here, we further develop this methodology on hydrogen bonded clusters, thus enabling us to relate the correlation pattern to the charge state of the cluster. 

The measurements were carried out at the soft X-ray beamline I411, MAX-lab, Lund. The design of our multi-ion coincidence 3D momentum imaging time-of-flight (TOF) spectrometer is described elsewhere.\cite{Laksman2013-spectrometer} The event trigger was provided by electron detection, and care was taken to measure only within a clean `coincidence regime', where false coincidences are kept to a minimum. The spectrometer axis was oriented perpendicular to both the X-ray polarization direction and the X-ray beam. The cluster jet was parallel to the polarization, and produced from a continuous supersonic nozzle expansion of gaseous ammonia (Linde, $<0.001$ permille impurity). We used a conical nozzle (20$\degree$ full opening angle, 150 $\mu$m throat) operated at a temperature of -10$\degree$C. In the following analysis, the longitudinal jet velocity was assumed constant (about 900 m/s) and the resulting momentum was subtracted from all measured particles. We have compared the response of the uncondensed molecular beam to those measured in effusive jet ammonia measurements, and verified that the molecular beam does not introduce artifacts in our analysis.

In order to estimate and tune the cluster size distribution, we used the detection of the stable dication, at a reported size (or number of molecules) $n=51$ (\ce{(NH_3)_51 H^{2+}_2}),\cite{Shukla1984, COOLBAUGH198919} in our experiments observed at $n=53$, Figure \ref{fig:1D_m2q}c. The average cluster size of ammonia clusters was estimated following an adiabatic expansion by the formalism of Bobbert et al,\cite{Bobbert_2002} which adapts the $\Gamma^*$ formalism of atomic clusters \cite{Hagena_1972, Karnbach_1993} to molecular clusters. The theorem predicts an average cluster size around 51 at a stagnation pressure between 0.1 and 0.4 bar.  Within this pressure range, the final optimization was performed by maximizing the measured dication signal at a constant photon energy of 420 eV. A typical mass spectrum of cluster fragments is shown in Figure \ref{fig:1D_m2q}b, and the molecular fragments in Figure \ref{fig:1D_m2q}a as a reference. The maximum intensity was found at a stagnation pressure of around 0.2-0.3 bar. In the following, data with different stagnation pressures (0.2 and 0.3 bar), and at photon energies above and below the 1s core hole (360 and 420 eV)  are merged, since the analysis does not reveal a statistically significant change between measurements.

The majority of events only contain the detection of one charged fragment, even when the expected charge state is higher (N1s ionization $\rightarrow$ Auger decay). This is partly a result of the incomplete detection of events (`aborted events'), due to a finite detection efficiency (of around 35\%). An improved model of Simon et al. \cite{Simon_1991} was applied to the data, which also includes a mass-dependent detection efficiency.\cite{Gilmore_2000} This model estimates the real number of charged fragments from the measured number of fragments (`single coincidence' for one, `double coincidence' for two fragments, etc.).
We attribute the measurement of non-aborted single coincidence events to mass-asymmetric fission \cite{Wu_2011}, resulting in measuring a small charged fragment correlated to an undetected larger cluster. No single coincidence is therefore analyzed further, due to the anticipated kinematic incompleteness.\cite{Geiger_2002}

In this work, we are interested in the cluster break-up into multiple charged fragments and their correlation, and we therefore discuss the double and triple coincidence events.

\begin{figure*}[t!]
\centerline{\includegraphics[width=0.9\textwidth]{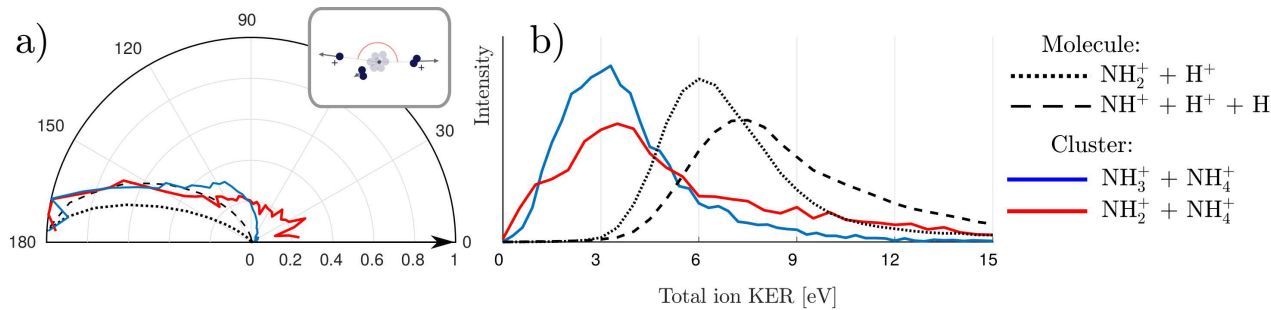}}
\caption{The kinetics of charged ammonia (NH$_3^+$, red) and amidogen (NH$_2^+$, blue) fragments, in coincidence with a single protonated cluster. The histogram in (a) shows the angular correlation between the momenta of both fragments (see inset), and the total kinetic energy release distribution (b). Both are compared to molecular correlations. The polar histogram in (a) shows the solid-angle corrected angle \cite{Hoener_2008} between fragments with one of the two fragment momenta aligned along the arrow. The molecular signals originate from \ce{NH_2+ + H^+} (black dotted) and \ce{NH_2+ + H^+ + H} (black dashed), upon 360 eV photoionization.}
\label{fig:unprot_mol}
\end{figure*}

At high photon energies, the majority of the observed cluster fragments were protonated (\ce{(NH_3)_n H^{+}}) due to chemistry within the multiply charged clusters. Coincidences between charged clusters and small charged molecular fragments (e.g. H$^+$, NH$^+$, etc) were not observed, except for weak channels containing non-protonated (NH$_3^+$) or amidogen (NH$_2^+$) molecules. Since the kinematics of fragmentation for weak and strong channels are different, we will discuss separately the breakup into unprotonated and protonated species.

\subsection*{Breakup into unprotonated fragments}



Pairs that contain either (NH$_3^+$) or (NH$_2^+$) have a channel strength of around 1\% and 3\%, respectively, based on the total double coincidence cluster signal. This branching ratio was found lower than that measured upon UV photoionization, which agrees with previous findings that show a decreasing amount of unprotonated fragments as photon energy increases.\cite{Shinohara_1984, Kaiser1991} Due to the high reactivity of ammonia (NH$_3^+$) or amidogen (NH$_2^+$), any interaction of these ions with surrounding molecules in the cluster will lead to their protonation,\cite{Peifer_1989} and we therefore expect that these molecular fragments broke out of the parent cluster almost freely. In such a case, the analysis of the kinematic of the pairs containing (NH$_3^+$) or (NH$_2^+$), and in particular the angular momenta correlation between fragments, should thus provide information about the fragmentation dynamics. In Figure \ref{fig:unprot_mol}a), the  mutual angles between momenta of molecular and cluster fragments are shown together with a two-body breakup (\ce{NH_2+ + H^+} in black dotted) and a three-body fragmentation (\ce{NH^+ + H^+ + H} in black dashed) as a reference. Note that in the latter case, a clear increase in distribution broadness of around 180\degree is observed for the three-body channel, despite the low mass of the neutral body (H). The momentum correlation of unprotonated fragments (\ce{NH_2+ + NH_4^+}, red) and (\ce{NH_3+ + NH_4^+}, blue) shows the same pattern. The (\ce{NH_2+ + NH_4^+}) channel has a low strength, and therefore shows non-zero intensity at small mutual angles, due to false coincidences. Overall, the angular momentum correlation of these fragments can best be compared to that of the three body molecular channel, which indicates that a third fragment has been produced. Furthermore, the strong anti-correlation of momenta is evidence that a relatively small momentum is imparted to the undetected fragment, suggesting it is a neutral molecule (no Coulomb repulsion). These findings support that the NH$_3^+$ and NH$_2^+$ fragments interact very little with other molecules, and points to the possibility that the charge sites could be located at the surface of the doubly-charged parent cluster.

Since the angular momentum correlation of unprotonated fragment pairs exhibits the pattern of a few-body breakup, the summed kinetic energy of both fragments yields the total Kinetic Energy Release (KER) of the fragmentation, where the kinetic energy of the undetected low-momentum particle is neglected. The latter is shown in Figure \ref{fig:unprot_mol}b, where the KER from molecular fragments are included as a reference as well. 
The KER of pairs containing \ce{NH_2+} and \ce{NH_3+} have a most probable value of around 3.5 eV. This KER is smaller than that from the molecular fragments, mainly due to the larger charge separation distance (CSD) upon breakup, and can be compared to two extreme cases. In one case, the charge transfer in the cluster is faster than nuclear motion, and the charge separation equals the equilibrium \ce{N-N} distance (3.37 \AA \cite{Tomoda1986}) of the neutral dimer, resulting in a Coulomb potential energy of 4.3 eV. In the alternative case of slow charge transfer, the \ce{N-N} distance equals that of the singly charged protonated dimer (1.57 \AA \cite{Wang_2002}), with 9.2 eV Coulomb energy (in both cases, the vacuum permittivity is assumed).
The measured KER is closer to the energy expected from the neutral distance than the contracted \ce{N-N} distance of the charged dimer, indicating that the dissociation process is faster than the typical \ce{N-N} nuclear relaxation time (the dimer does not reach equilibrium \ce{N-N} distance before dissociation). The 0.8 eV deviation towards lower observed KER can be explained by transfer of potential energy to rovibrational degrees of freedom,\cite{Jahnke2010, Vendrell_2010} or transfer to kinetic energy of an undetected neutral particle.

\begin{figure*}[t!]
\centerline{\includegraphics[width=0.7\textwidth]{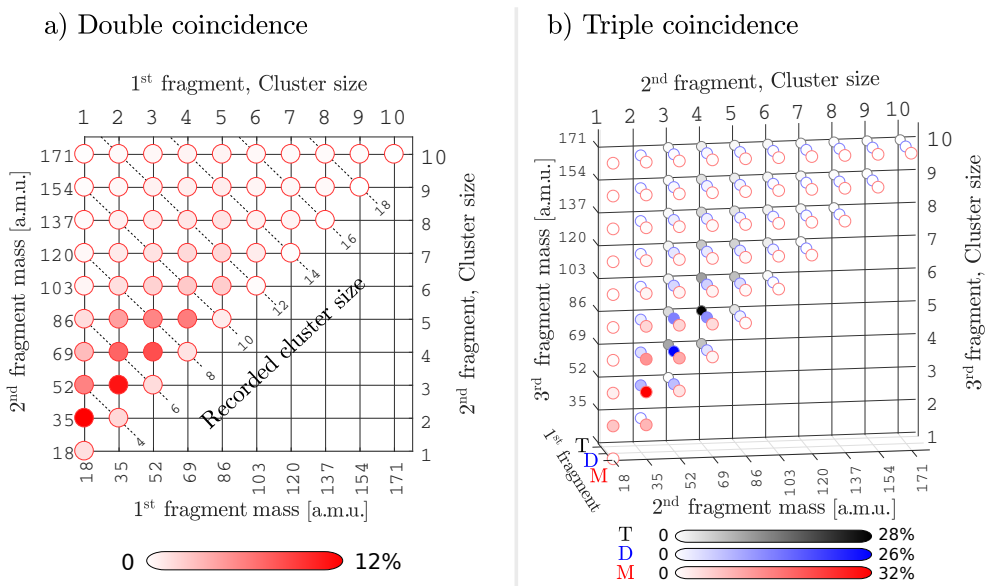}}
\caption{Relative branching ratios of protonated ammonia clusters ((\ce{NH_3})\ce{_n H+}) after photo-ionization (Photo-Electron, Photo-Ion Coincidence, PEPICO) in double coincidence (a,  PE(PI)\textsuperscript{2}CO)), and triple coincidence (b, PE(PI)\textsuperscript{3}CO). In the PE(PI)\textsuperscript{3}CO maps, events with a protonated monomer (M, red), dimer (D, blue) and trimer (T, black) as first fragment are selected. The `recorded cluster size' is indicated, defined as the total number of molecules observed in the event. The maps are separately normalized to highlight the most probable pair. The relative branching ratio is represented by the color intensity of the circles. The branching ratios for pairs of identical fragments are underestimated due to the detector dead time ($\sim 20$ ns)}
\label{fig:PEPICO}
\end{figure*}

Paradoxically, our observation leads to the conclusion that there is a slow proton transfer mechanism for these fragments, while the production of two separated charges upon ionization in the vicinity of the N1s edge requires dynamics faster than structural changes in the cluster. Without the coincident detection of electron kinetic energy, we can only establish the most plausible scenarios for the ultrafast dynamics triggered by ionization. 
It is theoretically predicted \cite{Kryzhevoi_2011} that the first repulsive electronic state with two single charges residing on neighboring molecules, \ce{NH_3^+ (2a_1^{-1})} - \ce{NH_3^+ (2a_1^{-1})}, lie at 24.7 eV. 
When the photon energy is below the N1s edge, photoionization of the 2a${_1}$ molecular orbital (26.8 eV \cite{Hjelte_2003}) may lead, via auto-ionization, to repulsive states that are energy-allowed. 
Direct double ionization can also occur, leading to the production of two holes (2e$^{-2}$, 2e$^{-1}$ 3$a_1^{-1}$ or 3a$_1^{-2}$) on a single molecule, and may transfer a proton to a neighboring site, within typically 200 fs.\cite{Farmanara_1999}
In both cases, two neighboring molecules will undergo Coulombic fragmentation with a KER that almost reflects the nitrogen-nitrogen distance between two neutral molecules, in line with our observation.
Above the N1s edge, we mostly produce two-hole states localized on one molecule.\cite{Lindblad_2008} However, we do not observe any photon energy dependence in the dissociation dynamics with the present statistics, indicating that the same efficient charge separation mechanism dominates below and above the N1s edge. 
Further charge transfer in a hydrogen bonded system implies a complex nuclear and structural rearrangement.\cite{agmon1995, Marx_2006} For the fastest process related to the proton as a charge carrier, it was shown that for a coordination number lower than 3, the proton motion requires larger structural changes,\cite{Meuwly_2001} slowing down the proton hopping mechanism. Since the average hydrogen bond coordination number on a cluster surface is lower (between 1 and 4 \cite{Beu_2001}) than the average coordination number in bulk ammonia (6, 3 donor and 3 acceptor bonds \cite{Okotrub1996}), the localization of charges at the surface would lead in some cases to dissociation via Coulomb explosion before proton hopping, resulting in the formation of unprotonated fragments. We can therefore speculate that the unprotonated fragment pairs are a result of ionization at the surface.

\subsection*{Breakup into protonated fragments}

 The distribution of protonated fragments of double coincidence is shown in Figure \ref{fig:PEPICO}a).
The channel strength of fragments is represented in the color intensity of the circles. 
We show triple coincidence events where the first fragment is a monomer (red), dimer (blue) or trimer (black) in Figure \ref{fig:PEPICO}b).
We find an increase of average charge state of clusters, compared to the isolated molecule, and assign this increase to ICD\cite{Kryzhevoi_2011} and intra-cluster electron scattering. Using the electron scattering cross sections,\cite{Zecca_1992, Jain_1988} we estimate that at most 30-40\% of 300 eV free electrons scatter, when emitted from the center of a cluster of 52 molecules in a shell structured cluster,\cite{Nakai2000201} with a near equal probability between ionization or elastic scattering.\cite{Zecca_1992}

We observe that most protonated cluster fragments appear at or below 5 molecules, the size where the ionic cluster can form the stable pentamer structure.\cite{Peifer_1989} The stability of the pentamer structure is specific to the ion, rather than the neutral,\cite{Shinohara_1985} which reveals molecular rearrangement during break-up.
It is established that delayed unimolecular evaporation in weakly bound clusters can occur on a microsecond timescale, i.e. during the flight time through the spectrometer, thus influencing the observed distribution of (non evaporated) parent fragments with respect to (metastable evaporated) daughter fragments. The importance of this effect can be estimated by the ratio between unimolecular evaporation lifetime $\tau_\text{m}$ and the nominal acceleration time of the ion in the electrostatic acceleration region $\tau_\text{acc}$ \cite{Gisselbrecht_2008}. In ammonia clusters, a unimolecular evaporation lifetime of $\tau_\text{m}$ from 8 to \SI{20}{\micro\second} for cluster sizes of 4 to 17 molecules can be determined from multi-photon UV absorption studies  \cite{Wei_1990}, while the acceleration time of $\tau_{acc}$ is about 2 to \SI{4}{\micro\second} in our experiment is a factor of four smaller than the typical evaporation lifetime. Hence, we can conclude that the majority of the measured intensity in the time-of-flight spectra represents the fragment mass directly after breakup, and only a few percent of the fragments have undergone unimolecular evaporation. 
In the following, we can introduce the `recorded cluster size' as the sum of the fragment cluster sizes before unimolecular evaporation has occurred (see Figure \ref{fig:PEPICO}a). The possible difference in produced and observed sizes is due to neutral fragment formation during break-up, as is predicted by classical molecular dynamics calculations on Morse clusters,\cite{Last_2002} or the fast `boiling off' of monomers \cite{Shinohara_1985} before break-up.

\begin{figure*}[t!]
\centerline{\includegraphics[width=0.7\textwidth]{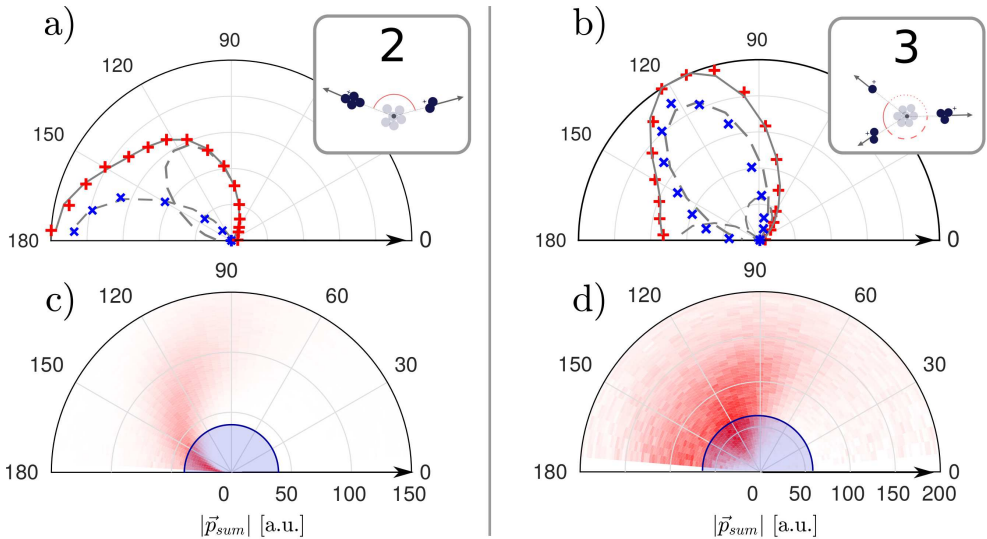}}
\caption{Fragment momentum angular correlation of clusters in double (a, c) and triple (b, d) coincidences. The histograms show the solid-angle corrected angle \cite{Hoener_2008} between fragments with one of the two fragment momenta aligned along the arrow. 
The histogram of mutual angles between fragment momenta (red plus, a,b) is fitted by the sum of Gaussian peaks (gray lines) at 180, 120 and 109.5 degrees (dashed).
The same momentum angle as in (a) and (b) are shown in (c) and (d) respectively, but now sorted in the momentum sum norm ($|\vec{p}_{\text{sum}}|$). 
The blue crosses (a, b) and blue region (c, d) show histograms of momentum-selected events ($|\vec{p}_{\text{sum}}| < $ 40 [a.u.] (two-body) and 60 [a.u.] (three-body).}
\label{fig:p_corr}
\end{figure*}

The expected fission-like character of the breakup can be verified by the momentum correlation of the fragments.\cite{Hoener_2008, Last_2002}
The cluster fragment momentum correlation, including all recorded cluster sizes, are presented in Figure \ref{fig:p_corr}. The double and triple coincidence events are shown on the left (a, c) and right (b, d), respectively. The direction of one of two momenta is directed along zero degrees (arrow), and the relative angle to the other momentum vector is plotted in a polar histogram as red crosses in Figure \ref{fig:p_corr} (a, b). In the case of triple coincidence, all combinations of two of the three momenta yield the same histogram, which are therefore presented in one curve here. The mutual angles range from almost perpendicular to a back-to-back formation.

In a multi-charged cluster, the charges will experience Coulomb interaction that will drive charges apart along minima in the potential energy surface, that maximize the mutual charge-to-charge distances.\cite{Casero_1988} The expected relative angles between their momenta are 180$\degree$ (2 charged fragments), 120$\degree$ (3 charged fragments, in plane) and 109.5$\degree$ (4 charged fragments, tetrahedron).
The observed distributions in Figure \ref{fig:p_corr} of two (a) and three (b) charged particles show high intensities at 180 and 120 degrees, respectively, but not exclusively at those angles due to the influence of higher-charged states. 
For example, the distribution of double coincidences (Figure \ref{fig:p_corr}a, red crosses) partly originates from triply charged cluster signal, where one of the charged fragments is undetected (a so-called `aborted event'.\cite{Simon_1991})
In order to account for the multiple contributions to the mutual angle histogram, we perform a multi-peak Gaussian fit centered at the characteristic angles (180\degree, 120\degree, etc.) in Figure \ref{fig:p_corr} (a)(gray dashed lines). The resulting total fit (gray solid line) shows an excellent agreement to the measured histogram, and the ratio between complete and aborted events determined from the fit matches the ratio calculated from the detection efficiency model (see method section). Consequently, we can confirm that the mutual angles of the detected fragments are governed by the number of charges in the cluster.

A detailed study of the fission mechanism requires the determination of the kinematics of all fragments. In order to restrict our analysis to `complete' detections, i.e. without undetected neutral particles or large undetected ionic clusters, we investigate only events where the residual momentum of undetected particles, determined by the momentum conservation as $|\vec{p}_{\text{sum}}|$, is negligible in comparison to the momenta of individual fragments. In the following, we select a subset of data by setting the value of $|\vec{p}_{\text{sum}}|$ to 40 and 60 a.u. for double and triple coincidence, respectively (blue regions in Figure \ref{fig:p_corr}(c, d)). This filter yields distributions (blue cross in Figure \ref{fig:p_corr}(a, b)) around the expected mutual angles (180\degree and 120 \degree) that are also found from the distribution fit, showing that in this data subset, the number of charges in the parent cluster is equal to the number of fragments detected.

\begin{figure*}[t!]
	\centering
	\centerline{\includegraphics[width=0.85\textwidth]{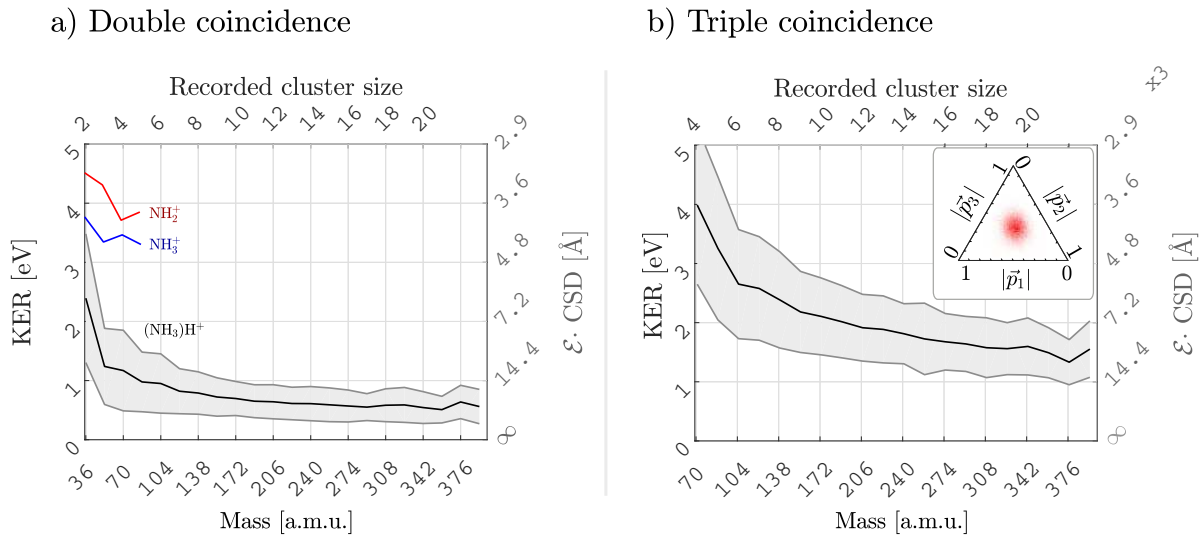}}
    \caption{The average KER (black solid line) and the standard deviations (gray zone) of protonated clusters after momentum-filtering ($|\vec{p}_{\text{sum}}| < $ 40, 60 [a.m.u.] for double (a) and triple (b) coincidence, respectively). The average KER of the channels containing a charged ammonia (blue) and amidogen (red) are shown as a reference. The inset in (b) shows the (normalized) momentum norm distribution of the triple coincidence momentum-filtered pairs in a Dalitz representation.\cite{Dalitz_1954} The normalization is done by the sum of all three momentum norms. The average KER is calculated over a range of 0-10 eV.}
\label{fig:KER}
\end{figure*}

In our experiments, the recorded cluster sizes are around five times smaller than the critical size of doubly and triply \cite{Kreisle_1986} charged clusters. In terms of the Liquid Droplet Model, this suggests that no energy barrier exists to prevent direct breakup (fissility parameter above 1). However, the clear anisotropy implies that the fragmentation is described by fission rather than explosion.\cite{Last_2002} Fission requires an energy barrier for direct breakup, resulting in delayed fragmentation.\cite{Casero_1988} The parent cluster size is thus most likely larger than the observed cluster size, resulting in a slower fragmentation and allowing charge separation to occur prior to breakup.
Charge transfer can be initiated by proton transfer upon ionization, which occurs in a few hundred femtoseconds.\cite{Farmanara_1999} After that, the charge transport to next neighbors involves molecular rearrangement.\cite{Markovitch_2008, Meuwly_2001}
Two mechanisms often considered are: the `Grotthuss' proton hopping mechanism, where the charge moves by transferring protons along hydrogen bonds, and `vehicular' transport, where the charge does not move over different molecules, but instead remains on a single moving molecule.
Both mechanisms play a role in the charge transport, and the relative importance is sensitive to the local electrostatic field.\cite{Cao_2014}
For instance, it is found in liquid water that the directionality of the proton hopping increases at and above a field strength of 0.07 V/\AA, making the hopping mechanism dominant over the vehicular diffusion.
In the case of two neighboring charged molecules in an ammonia cluster ($r_{N-N} = 3.37$ \AA), the imposed electric field is in the order of 1.2 V/\AA, and therefore we expect the charge separation to be dominated by proton hopping in the bulk. Assuming a proton hopping time of about 1.5 ps (as in liquid water,\cite{Meiboom_1961}) we can determine that the typical transport time for a charge in our ammonia cluster is in the order of few tens of picoseconds, providing hence an estimate of the timescale for fission to occur.

An important aspect of the fragmentation dynamics is the sharing of the Coulomb potential to kinetic and internal energy.
While the charge separates, the decrease of the Coulomb potential can be transfered to rovibrational excitations of the molecules along the charge path. It is shown experimentally \cite{Jahnke2010} and theoretically \cite{Vendrell_2010} that rovibrational degrees of freedom can share the potential energy during a Coulomb explosion in the water dimer. 
The Coulomb potential that is not converted to rovibrational degrees of freedom is then transfered to kinetic energy of the charged fragments.  
In Figure \ref{fig:KER}, the measured KER of the protonated cluster fragments are presented as a function of recorded size for double (a) and triple (b) coincidence. 
In the data shown, we have selected the subset of events that is considered kinematically complete, i.e. with negligible $|\vec{p}_{\text{sum}}| < $ (blue regions in Figure \ref{fig:KER}c,d), such that the sum of the measured kinetic energies represents the total kinetic energy of the reaction.
In all cases, we find that the average KER is monotonically decreasing as the recorded size increases, with a maximum at the smallest recorded cluster size of a few eV. The smallest KER stabilizes above size 20, at around 0.8 $\pm$ 0.2 eV for doubly charged clusters and 1.4$\pm$ 0.2 eV for triply charged clusters.
The size dependent decrease of the KER can partially be caused by the increasing number of rovibrational degrees of freedom in increasingly larger molecular clusters. We could speculate that the conversion efficiency of Coulomb potential to vibrational and rotational degrees of freedom increases as more degrees of freedom are available at larger fragment sizes.
This could in turn increase the rate of unimolecular evaporation during the flight time, and will affect our measurement on the order of 10 meV for the sizes under study,\cite{Wei_1990} which is still two orders of magnitude smaller than the measured KER. 

In a Coulomb explosion picture, the measured kinetic energy released is traditionally interpreted as the Coulomb potential at the instance of break-up, assuming that the initial kinetic energy of the particles at breakup is negligible.\cite{Geiger_2002, Ruhl_2003} This allows the conversion of KER to charge separation distance (CSD). An estimate of the CSD is provided in Figure \ref{fig:KER}, where the relative permittivity ($\mathcal{E}$) is assumed a constant.
In the case of three singly-charged fragments (Figure \ref{fig:KER}b), we assume an equal CSD between all three charges of a concerted breakup (triangular arrangement), which is justified by the near-equal momentum sharing shown in the inset Dalitz \cite{Dalitz_1954} plot in Figure \ref{fig:KER}b.
The measured CSD in doubly and triply charged clusters is larger than the recorded cluster size can account for in a Liquid Droplet Model \cite{Wu_2011} with vacuum permittivity. To match the recorded cluster size to the measured CSD, a permittivity of around three has to be used. Obviously, the simple Coulomb model with a constant permittivity cannot account for the sharing of the Coulomb potential with intra-cluster rotations and vibrations \cite{Jahnke2010, Vendrell_2010} and the fact that ammonia is a polarizable molecule (i.e. $\mathcal{E}>1$) with a permanent dipole. To interpret our observation, a more detailed calculation, including conversion of the Coulomb energy into rovibrational excitation, charge screening and proton transport is needed.
The latter could explain the increase in measured CSD from unprotonated to protonated fragments, as seen in Figure \ref{fig:KER}a. Indeed, a more efficient proton transport mechanism would allow to distinguish processes involving molecules with high hydrogen bond coordination numbers (bulk) from those at lower coordination numbers (surface).

In summary, by studying the dynamics of unstable multi-charged clusters, we found that the recorded sizes are smaller than the expected size distribution in the cluster beam. The existence of a potential energy barrier will allow time for charge transfer and transport to occur, leading to strong angular correlation of fragment momenta. 
In line with simulations on Morse clusters,\cite{Last_2002} one can speculate that in some cases the cluster elongates due to Coulomb repulsion. The two/three formed charges migrate, resulting in separate protonated cations (NH$_4^+$), acting as a seed for the formation of charged fragments, which eventually break out of a neutral host that has a negligible momentum. 
This could justify the large charge separation distance, compared to the predicted center-to-center distance. Our work is compatible with previous studies on large stable multi-charged clusters that host two/three separate singly charged ammonia cations within the bulk of neutral molecules.\cite{COOLBAUGH198919, Peifer_1989} In that work, the authors conclude that the cations are acting separately within the bulk neutral cluster and separate as far apart as possible through Coulomb repulsion.

\subsection*{Conclusion}

The investigation of unstable multiply charged ammonia clusters upon single photon multi-ionization allows the determination of the kinematics of the cluster cation fragmentation channels. While synchrotron radiation allows state-selective ionization, the absence of photon energy dependence points out the existence of an ultrafast charge separation process in the cluster, which is most likely related to proton transfer, followed by a fast statistical redistribution of the excess energy. During the time between ionization and breakup, typically on the order of a few tens of picoseconds, there is possibility for structural reorganization, allowing for proton hopping. The lack of proton mobility for molecules at the surface could be at the origin of the production of unprotonated fragments, such as ammonia or amidogen cations. More detailed information on these ultrafast processes (<10 ps) could be obtained experimentally by pump-probe measurements. Alternatively, multi-scale molecular dynamics simulations could be used to uncover the details of proton transfer mechanisms, such as improved estimates of timescales, the relative importance of vehicular and Grotthuss mechanisms, and the differences in surface and bulk ionizations in the charge transfer process after multi-ionization. We believe that the  electric field induced upon ionization allows for directionality of the proton motion, and can provide key insight into charge/proton transfer occurring at the nanoscale in electrolyte solution with potential applications in chemistry and biology.

\subsection*{Acknowledgements}
We would like to thank Olle Bj\"orneholm, Mikko-Heikki Mikkel{\"a}, Anna Sankari and Maxim Tchaplyguine for their support during the experiments, and Sylvain Maclot for critical reading of the manuscript. The research was supported by the Swedish Research Council, Crafoord and the Walter Gyllenberg foundation.\\
The final version of this manuscript can be found in published form: \url{DOI: 10.1039/c7cp06688k}




\bibliography{bibliography} 
\bibliographystyle{rsc} 

\end{document}